\documentclass[12pt,a4j,notitlepage,fleqn]{article}

\usepackage{setspace}
\usepackage[margin=1in]{geometry}
\usepackage[symbol]{footmisc}
\usepackage[T1]{fontenc}
\usepackage[utf8]{inputenc}
\usepackage{authblk}

\makeatletter

\newcommand{\fboxsubsec}[1]{
	\begin{flushleft}
		#1
	\end{flushleft}
	}
\newcommand{\fboxsubsubsec}[1]{
	\begin{flushleft}
		#1
	\end{flushleft}
	}
\renewcommand{\subsection}{\@startsection{subsection}{2}{0pt}
	{1ex}
	{0.5ex}
	{\reset@font\it\fboxsubsec}
	}
\renewcommand{\subsubsection}{\@startsection{subsubsection}{2}{0pt}
	{1ex}
	{0.5ex}
	{\reset@font\fboxsubsubsec}
	}
\makeatother

\title{Time-Varying Comovement of Foreign Exchange Markets}%

\author{Mikio Ito$^{a}$, \ Akihiko Noda$^{b,c}$\thanks{\scriptsize Corresponding Author. E-mail: noda@cc.kyoto-su.ac.jp, Tel. +81-75-705-1510, Fax. +81-75-705-3227.} \ and \ Tatsuma Wada$^{d}$

{\scriptsize ${}^{a}$ \it Faculty of Economics, Keio University, 2-15-45 Mita, Minato-ku, Tokyo 108-8345, Japan}

{\scriptsize ${}^{b}$ \it Faculty of Economics, Kyoto Sangyo University, Motoyama, Kamigamo, Kita-ku, Kyoto 603-8555, Japan}

{\scriptsize ${}^{c}$ \it Keio Economic Observatory, Keio University, 2-15-45 Mita, Minato-ku, Tokyo 108-8345, Japan}

{\scriptsize ${}^{d}$ \it Faculty of Policy Management, Keio University, 5322 Endo, Fujisawa, Kanagawa, 252-0882, Japan}}

\date{This Version: \today}

\renewcommand\thefootnote{\arabic{footnote}}

\usepackage[dvips]{graphicx}

\usepackage[]{natbib}%
\usepackage{amsmath,amssymb}%
\usepackage{ascmac}%
\usepackage{multirow}%
\usepackage{lscape}%
\usepackage{subfigmat}

\usepackage{pifont}%
\usepackage{arydshln}%
\usepackage[format=hang]{caption}
\usepackage[all]{xy}
\usepackage{url}
\bibpunct{(}{)}{;}{a}{}{,}

\def\hsymbu#1{\smash{\lower1.7ex\hbox{\huge$#1$}}}

\newcommand{\citetapos}[1]{\citeauthor{#1}'s \citeyearpar{#1}}

\begin{document}
\begin{titlepage}

\renewcommand{\thepage}{}
\renewcommand{\thefootnote}{\fnsymbol{footnote}}

\maketitle

\vspace{-10mm}

\noindent
\hrulefill

\noindent
{\bfseries Abstract:} A time-varying cointegration model for foreign exchange rates is presented. Unlike previous studies, we allow the loading matrix in the vector error correction (VEC) model to be varying over time. Because the loading matrix in the VEC model is associated with the speed at which deviations from the long-run relationship disappear, we propose a new degree of market comovement\ based on the time-varying loading matrix to measure the strength or robustness of the long-run relationship over time. Since exchange rates are determined by macrovariables, cointegration among exchange rates implies those macroeconomic variables share common stochastic trends. Therefore, the proposed degree measures the degree of market comovement. Our main finding is that the market comovement has become stronger over the past quarter century, but the rate at which market comovement strengthens is decreasing with two major turning points: one in 1995 and the other one in 2008. \\

\noindent
{\bfseries Keywords:} Foreign Exchange Markets; Market Comovement; Time-Varying Vector Error Correction Model.\\

\noindent
{\bfseries JEL Classification Numbers:} F31; F36; G14.
 
\noindent
\hrulefill

\end{titlepage}

\bibliographystyle{asa}%

\pagebreak

\section{Introduction}\label{ce_sec1}

It is well understood among researchers that the exchange rate dynamics is highly complex and that there is virtually no consensus as to what econometrics model best describes the time-series process of exchange rates. While many exchange rates are known to have unit roots, finding cointegrating relationship in several exchange rates has attracted relatively little interest with some exceptions such as \citet{baillie1989cst}. Because an exchange rate cointegrating with another exchange rate means they share a common stochastic trend, discovering cointegration in exchange rates is tantamount to discovering common trends therein. As \citet{baillie1989cst} argue, many of exchange rates are found to have a unit root or stochastic trend. If those exchange rates are not cointegrating, then any shock to an exchange rate has a permanent effect, irrespective of other exchange rates thereby diverging from them. This situation is somewhat awkward because it is hard to imagine that interest rate differentials, one of main determinants of exchange rate dynamics, do not exhibit international comovement. Yet, a skeptical view on cointegrating exchange rates is presented by, for example, \citet{diebold1994oce}, who claim that the martingale model is better at out-of-sample forecast than the cointegrating model. They also find weaker evidence of cointegration in exchange rates than do \citet{baillie1989cst}. As explained in \citet{engle2007erm}, an important implication of \citet{engle2005erf} is that a large discount factor could obscure the cointegrating relationship in exchange rates when the exchange rates are generated by the present value model, even when the exchange rates are indeed cointegrated. 

All in all, whether several exchange rate are cointegrated is rather an empirical question. However, investigating whether the cointegrating relationship is stable over time is more meaningful in economics because exchange rate dynamics may not be described by a single model, or because unstable relationship could be reflective of extraordinary events that alter the global financial environment. Also possible is that the global financial markets are constantly changing, and hence, the relationship between exchange rates too are constantly changing. 

This very idea of the time-varying market environment or market integration is in line with \citet{ito2014ism} who show that the global stock markets become, from time to time, efficient, in \citetapos{fama1970ecm} sense, with varying degrees. As the main contribution of this paper, we propose the degree of market comovement and then, we show that the degree of market comovement has monotonically increased over time in the past quarter century. However, the increase in the degree of market comovement was found to be at a diminishig pace, suggesting that the cointegrating relationship may have a ceiling that forbids the cointegrating relationship from strengthening any further.      

This paper is organized as follows. In Section \ref{ce_sec2}, we present our error correction model that allows some key parameters to be time-varying. In the same section, we propose a new measure for foreign exchange markets, namely, the degree of market comovement. The exchange rate data together with preliminary unit root test results, are given in Section \ref{ce_sec3}. Section \ref{ce_sec4} provides the main results and some discussion on the time-varying nature of the global exchange market. Our conclusion is in Section \ref{ce_sec5}.

\section{Model}\label{ce_sec2}

This section presents our method to capture the time-varying nature of foreign exchange markets. The main building block of our model is a vector error correction (VEC) model, which supposes that there are cointegrating relationships or long-run relationships among the variables in our model. In particular, our idea stems from the fact that the VEC model elucidates the adjustment process to the long-run relationships. As the following subsections explain, the crux of our model is that the adjustment process or the speed of adjustment to the long-run relationships can vary over time, reflecting the changing environment in the global foreign exchange markets.

\subsection{Exchange Rate Dynamics and Cointegration}

Let us suppose the natural log of the spot Japanese yen per U.S. dollar exchange rate $s_{t}^{J}$ and the natural log of the spot Canadian dollar per U.S. dollar $s_{t}^{C}$ are cointegrated. Then, there is an error correction model representation: 
\begin{eqnarray*}
\Delta s_{t}^{J} &=&\alpha_{1}\left( s_{t-1}^{J}-\beta_{1}s_{t-1}^{C}\right) +\gamma_{1,1}\Delta s_{t-1}^{J}+\gamma_{1,2}\Delta s_{t-1}^{C}+e_{1t}, \\
\Delta s_{t}^{C} &=&\alpha_{2}\left( s_{t-1}^{J}-\beta_{1}s_{t-1}^{C}\right) +\gamma_{2,1}\Delta s_{t-1}^{J}+\gamma_{2,2}\Delta s_{t-1}^{C}+e_{2t},
\end{eqnarray*}
where $\beta =\left( 1,\beta _{1}\right) $ is called the cointegrating vector and $s_{t-1}^{J}-\beta _{1}s_{t-1}^{C}=0$ is called the long-run relationship, which is stationary. The coefficients on the long-run relationship, $\alpha_{1}$ and $\alpha_{2}$ are interpreted as speed at which any deviations from the long-run relationship disappear\footnote{More precisely, the coefficient $\alpha_{1}$ or $\alpha_{2}$ is associated with a half-life of the deviation from the cointegrating relationship.}. As we shall see in the following subsections, it is possible that the speed of adjustment changes over time, perhaps due to the fact that the environment of the international financial markets is changing. 

\subsection{The Vector Error Correction Model}

Assuming there are some cointegrating relationships, let us consider a vector error correction (VEC) model for $m$-vector time series $X_{t}$ 
\begin{equation}
\Delta X_{t}=\Gamma_{1}\Delta X_{t-1}+\cdots +\Gamma_{k}\Delta X_{t-k}+\Pi_{k}X_{t-k}+\boldsymbol\mu+\boldsymbol\varepsilon_{t}, \label{VEC}
\end{equation}%
where $\Delta X_{t}=X_{t}-X_{t-1}$, $\boldsymbol\mu$ is a vector of intercepts, and $\boldsymbol\varepsilon _{t}$ is a vector of error terms. 

The VEC model (\ref{VEC}) suggests that $\Delta X_{t}$ consists of a stationary part, $\Gamma_{1}\Delta X_{t-1}+\cdots+\Gamma_{k}\Delta X_{t-k}$, and an error correction term, $\Pi_{k}X_{t-k}$, which is also a stationary process when each variable in the vector $X_{t}$ is an integrated process of order one (often denoted as $I(1)$). Since we apply the VEC model to a set of variables whose first differences are stationary, $\Pi_{k}X_{t-k}$ is a vector of stationary process with a\ zero mean vector; and $\Pi_{k}X_{t-k}$ includes some long-run relationships among the variables in $X_{t}$. Note that the $m\times m$ matrix $\Pi_{k}$ is a singular matrix, i.e., the rank of $\Pi_{k}$ is $r$, which is less than $m$. Hence, if we decompose the matrix $\Pi_{k}$ such that $\Pi_{k}=\boldsymbol\alpha\boldsymbol\beta^{\prime }$, then $\boldsymbol\alpha$ is an $m\times r$ matrix; $\boldsymbol\beta^{\prime }$ is an $r\times m$; and both are rank-$r$ matrices. 

The long-run relationships among the variables in the system, or the cointegration is described as $\boldsymbol\beta^{\prime }X_{t-k}=0$, and hence, $\boldsymbol\beta$ is called the cointegrating matrix. We pay special attention to $\boldsymbol\alpha$, which indicates how quickly the exchange rates in the system restore the long-run relationships when deviations from such relationships occur. This is because a rapid adjustment implies strong or robust cointegrating relation, thus the determinants of exchange rate dynamics such as interest rate differentials, risk premia, and price level differentials should be following common stochastic trends. In this sense, the quicker the adjustment, the stronger is comovement. Furthermore, by allowing $\boldsymbol\alpha$ to be time-varying, we will capture the time-varying nature of market comovement by estimating the matrix changing over time with time-varying VEC model.

\subsection{The Time-Varying VEC Model}

It is not obvious whether the loading matrix $\boldsymbol\alpha$ is, in fact, time-varying. As an assessment, we apply \citetapos{hansen1992a} parameter constancy test to a VEC model. we impose the following parameter dynamics to a VEC model so that we can estimate the loading matrix changing over time. For given constant $\boldsymbol\beta$, we estimate $\Gamma$ and $\boldsymbol\alpha$ assuming 
\begin{eqnarray*}
\Gamma_{t} &=&\Gamma_{t-1}+u_{t} \\
\boldsymbol\alpha_{t} &=&\boldsymbol\alpha_{t-1}+v_{1t}
\end{eqnarray*}
It should be noted that simultaneous estimation of $\boldsymbol\alpha_{t}$ and $\boldsymbol\beta_{t}$ for each $t$ is infeasible due to an identification problem of a linear regression\footnote{See online appendix for a more detail discussion. It is available at \url{http://at-noda.com/appendix/efficiency_integration_appendix.pdf}.}.

\subsection{The Degree of Market Comovement}

When the loading matrix $\boldsymbol\alpha$ is not stable for the whole sample period, then it is reasonable to assume the speed of adjustment changes over time. Since the long-run relationship in exchange rates is presumably due to comovement in interest rate differentials, risk premia, or price level differentials, a rapid adjustment to the long-run relationship is associated with a strong comovement among those variables. For this, we propose the new measure of international comovement, called the degree of market comovement, which can be computed from the loading matrix $\boldsymbol\alpha$: 
\begin{equation*}
\zeta _{t}=\sqrt{\max\lambda\left(\boldsymbol\alpha_{t}\boldsymbol\alpha_{t}^{\prime }\right)}
\end{equation*}%
where $\max \lambda \left( A\right) $ is the largest eigenvalue of a matrix $A$. Note that the degree of market comovement is similar to the one proposed by \citet{ito2014ism} who examine international stock market efficiency and quantify the time-varying nature of market efficiency. The greater the degree of market comovement, the faster it is for foreign exchange markets to return to their long-run relationship when deviations from the cointegrating relationship arise.  

\subsection{Confirming Our Assumption of Constant Cointegrating Vectors}

Among the several assumptions we make about our model, one substantial assumption, namely, the cointegrating vectors being constant over time while the loading matrix being time-varying, needs to be justified. It is conceivable that the fact that the matrix $\Pi$ is time-varying, instead implies the number of cointegrating vectors (as well as cointegrating vectors themselves) are time varying. In order to confirm the stability of the cointegrating vectors, we apply the state-of-the-art econometrics test proposed by Qu (2007)\footnote{As detailed in \citet{juselius2006cvm}, there are a number of econometrics tests that investigate the constancy of cointegrating vectors. In our view, however, the \citet{qu2007scd} test is best suited for our purpose because it considers a variety of test statistics and their asymptotic properties, without assuming the timing of possible structural breaks.} to the exchange rate data. In essence, \citetapos{qu2007scd} test allows a researcher to assess whether the number of cointegrating vectors has changed during a subsample, say, the time between $T_{a}$ and $T_{b}$, where neither $T_{a}$ nor $T_{b}$ is known to the researcher. As the null hypothesis, \citetapos{qu2007scd} test states the number of cointegrating vectors is constant for the whole sample. Hence, not being able to reject the null hypothesis can be interpreted as justification of our assumption that the cointegrating vectors are stable over time.

\section{Data}\label{ce_sec3}

We utilize average monthly data on spot and forward exchange rates for three developed countries (Canada, Japan, and UK) from May 1990 to July 2015, which are taken from Thomson Reuter Datastream. For the forward exchange rates data, we use a nearby (one month) contract month following a number of the earlier studies. We take the natural log of the spot and forward exchange rates to obtain the level data, and we also take the first difference of the natural log of the level data to compute the returns on spot and forward exchange rates.
\begin{center}
(Table \ref{ce_table1} around here)
\end{center}
Table \ref{ce_table1} provides some descriptive statistics and the results of the ADF-GLS tests. For the unit root tests, the ADF-GLS test of \citet{elliott1996eta} is applied. We employ the modified Bayesian information criterion (MBIC) instead of the modified Akaike information criterion (MAIC) to select the optimal lag length. This is because we are unable to find the evidence of size-distortions (see \citet{elliott1996eta}; \citet{ng2001lls}) in the estimated coefficient of the detrended series, $\widehat{\psi}$. It is widely known that the logarithmic spot and forward exchange rates are both integrated of order one (or $I(1)$ process), so that the differences are stationary (or $I(0)$ process) variables.

\section{Empirical Results}\label{ce_sec4}

\subsection{Preliminaries}
It is our first order business to determine whether our data on exchange rates exhibit non-stationarity, more specifically, having a unit root. Shown in Table \ref{ce_table1}, the level data (upper panel) exhibit non-stationarity as the ADF-GLS test uniformly fails to reject the null hypothesis of the series possessing a unit root, while the same test is able to reject the null hypothesis once the first differences of each series is computed (lower panel).

Having confirmed that our data have unit roots, we then proceed to investigate whether the data have cointegrating relationships. To this end, we utilize \citetapos{johansen1988sac} maximum eigenvalue test and \citetapos{johansen1991eht} trace test.
\begin{center}
(Table \ref{ce_table2} around here)
\end{center}
Here, we assess the number of the cointegrating vectors in Table \ref{ce_table2}. The null hypothesis for both tests are presented in the first column of Table \ref{ce_table2}. From the first row we can conclude that the null hypothesis of no cointegrating vector is rejected at the 10\% level of significance by both the maximal eigenvalue test and the trace test. Yet, it is conclusive that the there are no more than one cointegrating vector. This is due to the fact that the maximal eigenvalue test whose alternative in the second row is 2 cointegrating vectors, cannot reject the null hypothesis and neither can the trace test whose alternative states there are more than 1 cointegrating vectors.
\begin{center}
(Table \ref{ce_table3} around here)
\end{center}
We report the estimates of our VEC models in Table 3, where each column corresponds to each VEC model showing coefficients on the error-correction term in the lower half of the table.

\subsection{The Time-Varying Model}

\begin{figure}[bp]
 \caption{Time-Varying Market Comovement and its Accelaration}\label{ce_fig1}
 \begin{center}
 \includegraphics[scale=0.5]{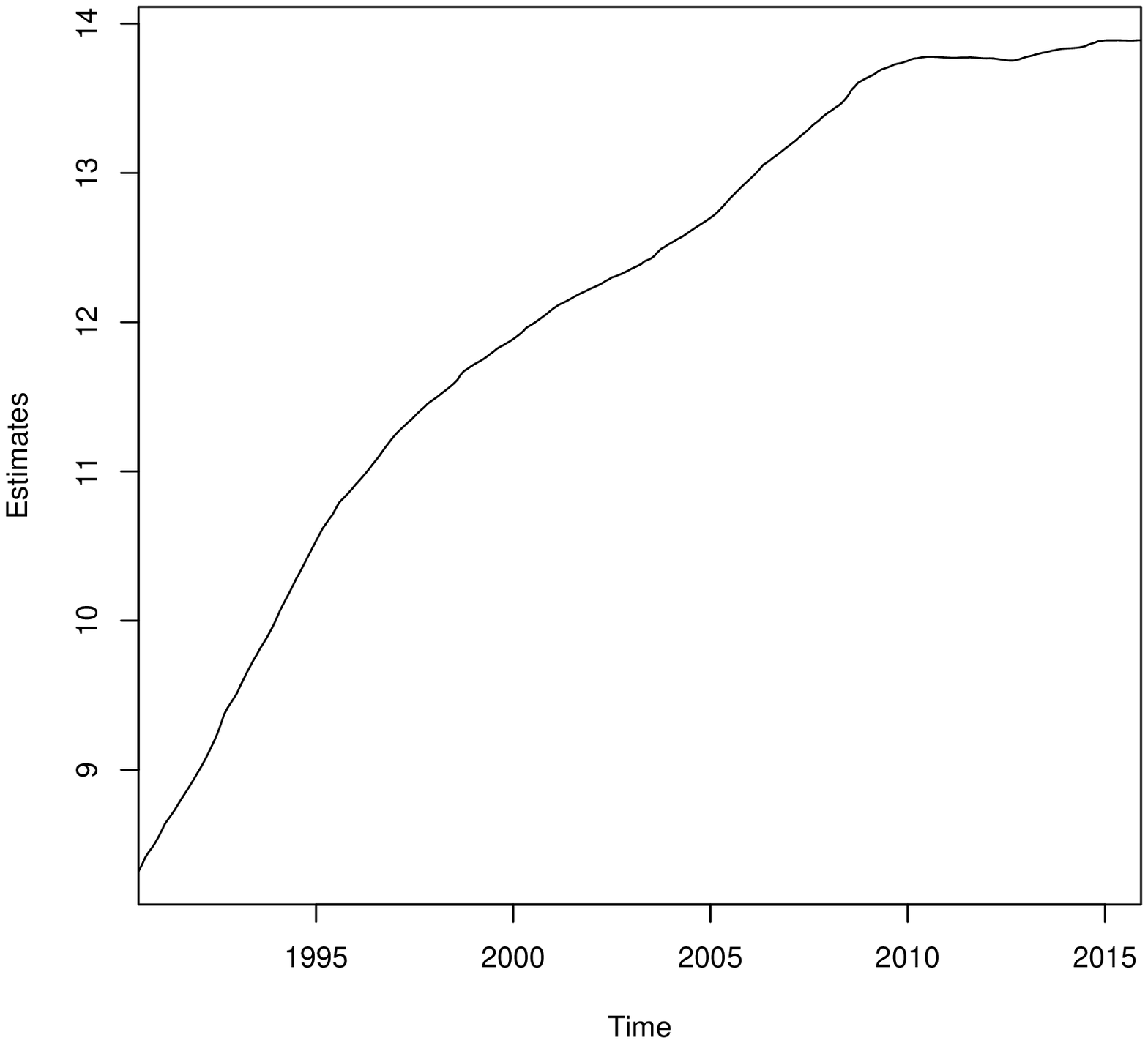}
 \includegraphics[scale=0.5]{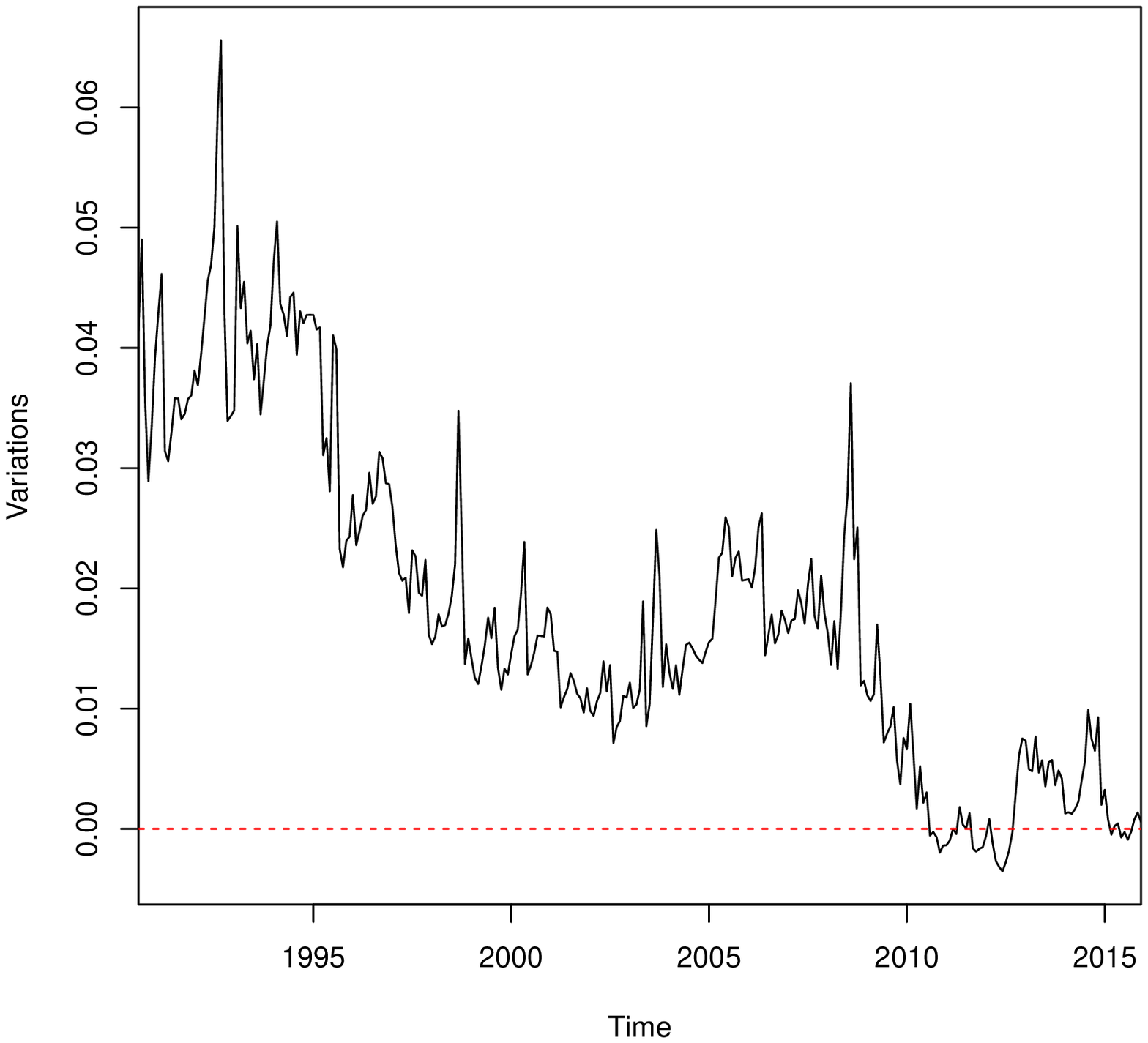}
\vspace*{3pt}
{
\begin{minipage}{350pt}
\footnotesize
\underline{Notes}: R version 3.3.1 was used to compute the estimates and the calculates.
\end{minipage}}%
\end{center}
\end{figure}

Up until now, we assume the VEC model has time-invariant parameters. Let us relax this assumption and use the time-varying parameter model that presumably better capture the dynamics of exchange rates, by taking into account constantly changing market environment. So, how different a picture can we get once we apply the time-varying VEC model? But first, we need to make sure whether the cointegrating relationship, or the cointegrating vector is preserved for the entire sample period. By the \citet{qu2007scd} test, we can conclude that the number of cointegrating vectors among the Japanese yen-U.S. dollar, Canadian dollar-U.S. dollar, and pound sterling-U.S. dollar exchange rates is constant over time at the 1\% level of significance. 
\begin{center}
(Table \ref{ce_table4} around here)
\end{center}
Instead of presenting the detailed estimates about our time-varying VEC model (that are available in the appendix), here we report the degree of market comovement $\zeta_{t}$. As we discuss in Section \ref{ce_sec2}, the larger $\zeta_{t}$, the quicker it is for the foreign exchange markets to adjust to their long-run relationship.
\begin{center}
(Figure \ref{ce_fig1} around here)
\end{center}
What the upper panel of Figure \ref{ce_fig1} shows is the foreign exchange markets that are increasingly rapidly restoring the long-run relationship when unanticipated shocks hit the markets. The lower panel provides a close look at the time-varying nature of the speed of adjustment, more precisely, the lower panel exhibits changes in the speed of adjustment. It is interesting to observe that the speed at which the markets restore the long run relationship has been increasing over time, but its rate of increase has diminished over the sample period. Notably, while an eyeball inspection, the changes in the rate of increase occurred at least twice, once in around 1995 and the second time in 2008. The second time of the change seemingly coincides with the global financial crisis, while the first time is likely associated with the Mexican peso crisis.

\section{Concluding Remarks}\label{ce_sec5}

A novel model for exchange rates, taking into account the time-varying nature of market environment, is presented. Our approach enables us to utilize cointegrating relationship among exchange rates, with the loading matrix in the VEC model changing over time. Since the loading matrix can be interpreted as the speed of adjustment or the strength of the cointegrating relationship in the exchange rates that are determined by various macroeconomic variables sharing common stochastic trends, we call the new degree derived from the loading matrix \textquotedblleft degree of market comovement.\textquotedblright\ With the new degree, we find that market
comovement has become stronger, but at a decreasing pace with two large turning points in 1995 and in 2008. 

As a new direction of research, one may move forward with the idea of time-varying cointegration by considering the exchange rate system that has cointegration relation at a time, but not the whole sample period. Those models can assess whether the claim made by \citet{engle2005erf} is empirically valid.

\section*{Acknowledgments}

We would like to thank conference participants at the 91th Annual Conference of the Western Economic Association International for their helpful comments and suggestions. We also acknowledge the financial assistance provided by the Japan Society for the Promotion of Science Grant in Aid for Scientific Research No.26380397 (Mikio Ito), No.15K03542 (Akihiko Noda), and No.15H06585 (Tatsuma Wada). All data and programs used for this paper are available on request.

\clearpage

\bibliography{paperdatabase}

\setcounter{table}{0}
\renewcommand{\thetable}{\arabic{table}}

\clearpage

\begin{table}[tbp]
\caption{Descriptive Statistics and Unit Root Tests}\label{ce_table1}
\begin{center}
\scriptsize
\begin{tabular}{llrrrrcccccc}\hline\hline
 &  & Mean & SD & Min & Max &  & ADF-GLS & Lags & $\hat{\phi}$ &  & $\mathcal{N}$ \\\hline
Level &  &  &  &  &  &  &  &  &  &  & \\
 & $S_{CA}$ & 0.2175  & 0.1452  & -0.0454  & 0.4697  &  & -1.2364  & 1 & 0.9948  &  & \multirow{6}*{308}\\
 & $F_{CA}$ & 0.2180  & 0.1449  & -0.0447  & 0.4699  &  & -1.2470  & 1 & 0.9947  &  & \\
 & $S_{JP}$ & 4.6918  & 0.1486  & 4.3396  & 5.0352  &  & -1.6915  & 1 & 0.9902  &  & \\
 & $F_{JP}$ & 4.6899  & 0.1483  & 4.3393  & 5.0344  &  & -1.6844  & 1 & 0.9903  &  & \\
 & $S_{UK}$ & -0.4965  & 0.0905  & -0.7279  & -0.3387  &  & -2.9910  & 1 & 0.9700  &  & \\
 & $F_{UK}$ & -0.4953  & 0.0900  & -0.7269  & -0.3373  &  & -2.9969  & 1 & 0.9698  &  & \\\hline
First Difference &  &  &  &  &  &  &  &  &  &  & \\
 & $\Delta S_{CA}$ & 0.0005  & 0.0167  & -0.0626  & 0.1083  &  & -11.8715  & 0 & 0.3648  &  & \multirow{6}*{307}\\
 & $\Delta F_{CA}$ & 0.0005  & 0.0167  & -0.0626  & 0.1077  &  & -11.8337  & 0 & 0.3676  &  & \\
 & $\Delta S_{JP}$ & -0.0007  & 0.0261  & -0.1095  & 0.0823  &  & -11.5044  & 0 & 0.3948  &  & \\
 & $\Delta F_{JP}$ & -0.0007  & 0.0261  & -0.1094  & 0.0831  &  & -11.5039  & 0 & 0.3948  &  & \\
 & $\Delta S_{UK}$ & 0.0004  & 0.0228  & -0.0583  & 0.1066  &  & -12.2763  & 0 & 0.3385  &  & \\
 & $\Delta F_{UK}$ & 0.0004  & 0.0227  & -0.0585  & 0.1080  &  & -12.2221  & 0 & 0.3424  &  & \\\hline\hline
\end{tabular}
{
\begin{minipage}{420pt}
\vspace*{3pt}\scriptsize
{\underline{Notes:}}
\begin{itemize}
\item[(1)] ``ADF-GLS'' denotes the ADF-GLS test statistics, ``Lags'' denotes the lag order selected by the MBIC, and ``$\hat\phi$'' denotes the coefficients vector in the GLS detrended series (see equation (6) in \citet{ng2001lls}).
\item[(2)] In computing the ADF-GLS test, a model with a time trend and a constant is assumed. The critical value at the 1\% significance level for the ADF-GLS test is ``$-3.42$''.
\item[(3)] ``$\mathcal{N}$'' denotes the number of observations.
\item[(4)] R version 3.3.1 was used to compute the statistics.
\end{itemize}
\end{minipage}}%
\end{center}
\end{table}

\clearpage

\begin{table}[tbp]
\caption{Johansen's Cointegration Tests}\label{ce_table2}
  \begin{center}
\begin{tabular}{llrrrrrrrrl}\hline\hline
 & & & & & \multicolumn{2}{c}{Maximal Eigenvalue} & & \multicolumn{2}{c}{Trace} & \\\cline{6-7}\cline{9-10}
 & & & Eigenvalues & & Test Stats & CV(10\%) & & Test Stats & CV(10\%)\\\hline
 & None & & 0.1352 & & 44.45 & 37.45 & & 100.19 & 97.18\\
 & At most 1 & & 0.0826 & & 26.39 & 31.66 & & 55.74 & 71.86\\\hline\hline
\end{tabular}
{
\begin{minipage}{400pt}
\vspace*{3pt}\footnotesize
{\underline{Notes:}}
\begin{itemize}
 \item[(1)] ``Maximal Eigenvalue'' and ``Trace'' denote the \citetapos{johansen1988sac} maximal eigenvalue test and \citetapos{johansen1991eht} trace test, respectively.
 \item[(2)] ``Test Stats'' and ``CV(10\%)'' denote the test statistics and the critical values at the 10\% significance level for the each tests, respectively.
 \item[(3)] R version 3.3.1 was used to compute the statistics.
\end{itemize}
\end{minipage}}%
  \end{center}
\end{table}

\clearpage

\begin{table}[tbp]
\caption{Time-Invariant VEC Model Estimations}\label{ce_table3}
\begin{center}
\begin{tabular}{cccccccc}\hline\hline
 &  & $S_{CA}$ & $F_{CA}$ & $S_{JP}$ & $F_{JP}$ & $S_{UK}$ & $F_{UK}$\\\hline
Difference &  &  &  &  &  &  & \\
 & \multirow{2}*{$S_{CA}$} & -0.4209  & -0.2370  & 6.0713  & 6.1933  & 2.0254  & 2.1560 \\
 &  & [1.8453]  & [1.8414]  & [4.5597]  & [4.5475]  & [4.4885]  & [4.4183] \\
 & \multirow{2}*{$F_{CA}$} & 0.7703  & 0.5885  & -6.0537  & -6.1739  & -1.6558  & -1.7854 \\
 &  & [1.8608]  & [1.8560]  & [4.5810]  & [4.5684]  & [4.4883]  & [4.4181] \\
 & \multirow{2}*{$S_{JP}$} & 1.6242  & 1.6664  & -0.7974  & -0.5514  & 1.0198  & 0.9973 \\
 &  & [1.8997]  & [1.8902]  & [3.3395]  & [3.3475]  & [3.0337]  & [3.0269] \\
 & \multirow{2}*{$F_{JP}$} & -1.6031  & -1.6460  & 1.1054  & 0.8580  & -0.9913  & -0.9692 \\
 &  & [1.9188]  & [1.9093]  & [3.3492]  & [3.3573]  & [3.0485]  & [3.0412] \\
 & \multirow{2}*{$S_{UK}$} & 0.1174  & 0.0667  & -3.9585  & -4.1291  & -6.0241  & -5.8113 \\
 &  & [2.1271]  & [2.1253]  & [3.1513]  & [3.1505]  & [3.5704]  & [3.5750] \\
 & \multirow{2}*{$F_{UK}$} & -0.1774  & -0.1263  & 3.9517  & 4.1228  & 6.2729  & 6.0556 \\
 &  & [2.1210]  & [2.1200]  & [3.1599]  & [3.1593]  & [3.6079]  & [3.6115] \\\hline
Level &  &  &  &  &  &  & \\
 & \multirow{2}*{Constant} & -0.0334  & -0.0334  & 0.2006  & 0.1995  & 0.1467  & 0.1452 \\
 &  & [0.0470]  & [0.0468]  & [0.0680]  & [0.0680]  & [0.0470]  & [0.0467] \\
 & \multirow{2}*{$S_{CA}$} & 0.8047  & 0.9465  & 3.3757  & 3.3233  & 2.2904  & 2.2845 \\
 &  & [1.1232]  & [1.1317]  & [2.6344]  & [2.6460]  & [2.2176]  & [2.2168] \\
 & \multirow{2}*{$F_{CA}$} & -0.8390  & -0.9808  & -3.3467  & -3.2952  & -2.2552  & -2.2494 \\
 &  & [1.1254]  & [1.1337]  & [2.6427]  & [2.6543]  & [2.2191]  & [2.2181] \\
 & \multirow{2}*{$S_{JP}$}& 1.3304  & 1.2953  & -1.0766  & -0.9938  & -1.5576  & -1.5626 \\
 &  & [1.0113]  & [1.0109]  & [1.3641]  & [1.3615]  & [1.1244]  & [1.1277] \\
 & \multirow{2}*{$F_{JP}$} & -1.3189  & -1.2837  & 1.0302  & 0.9477  & 1.5173  & 1.5226 \\
 &  & [1.0135]  & [1.0131]  & [1.3695]  & [1.3668]  & [1.1194]  & [1.1226] \\
 & \multirow{2}*{$S_{UK}$} & -2.0125  & -2.0557  & -1.4713  & -1.4990  & -1.8191  & -1.7604 \\
 &  & [1.5610]  & [1.5579]  & [2.2110]  & [2.2115]  & [2.3796]  & [2.3631] \\
 & \multirow{2}*{$F_{UK}$} & 2.0478  & 2.0910  & 1.4478  & 1.4764  & 1.7452  & 1.6862 \\
 &  & [1.5691]  & [1.5658]  & [2.2266]  & [2.2271]  & [2.3855]  & [2.3688] \\\hline
 $\bar{R}^2$ &  & 0.0984  & 0.1005  & 0.1032  & 0.1026  & 0.1918  & 0.1900 \\
 $L_C$ &  & \multicolumn{6}{c}{65.5233} \\\hline\hline
\end{tabular}
{
\begin{minipage}{400pt}
\vspace*{3pt}\footnotesize
{\underline{Notes:}}
\begin{itemize}
\item[(1)] ``${\bar{R}}^2$'' denotes the adjusted $R^2$, and ``$L_C$'' denotes \citetapos{hansen1992a} joint $L_C$ statistic with variance.
\item[(2)] \citetapos{newey1987sps} robust standard errors are in brackets.
\item[(3)] R version 3.3.1 was used to compute the estimates and the statistics.
\end{itemize}
\end{minipage}}%
\end{center}%
\end{table}%

\clearpage

\begin{table}[tbp]
\caption{Qu's Cointegration Order Change Tests}\label{ce_table4}
 \begin{center}
\begin{tabular}{cccccccc}\hline\hline
& & & $SupQ^1$ & $SupQ^2$ & $WQ$ & $SQ$ &\\\cline{4-7}
& Test Stats & & 8.55  & 10.81 & 8.55 & 15.94 &\\
& CV (1\%) & & 11.24  & 16.41 & 11.42 & 22.03 &\\\hline\hline
\end{tabular}
{
\begin{minipage}{250pt}
\vspace*{3pt}\scriptsize
{\underline{Notes:}}
\begin{itemize}
 \item[(1)] ``$SupQ^1$'' and ``$SupQ^2$'' denote the values allowing for one break and two breaks, respectively.
 \item[(2)] ``$WQ$'' and ``$SQ$'' also denote consistent test stats when we suppose less than three breaks based on the maximum and sum of $SupQ^1$ and $SupQ^2$, respectively.
 \item[(3)] R version 3.3.1 was used to compute the statistics.
\end{itemize}
\end{minipage}}%
\end{center}
\end{table}

\clearpage

\end{document}